\documentclass[a4paper,11pt]{article}
\usepackage{pos}
\usepackage[english]{babel}
\usepackage[utf8]{inputenc}
\usepackage[T1]{fontenc}
\usepackage{graphicx}
\usepackage{subfig} 
\title{First Results and Prospects for $\tau$ Lepton Physics at Belle~II}
\manuallySeparateAuthors
\author*[a]{Thomas Kraetzschmar}
\author{on behalf of the Belle II Collaboration}

\affiliation[a]{Max Planck Institute for Physics,\\
	Föhringer Ring 6,
	80805 Munich,
	Germany}

\emailAdd{kraetzsc@mpp.mpg.de}

\abstract{
Belle II has a broad $\tau$ physics program, in particular in searches for lepton flavour and lepton number violations. Belle~II profits from the relatively large  $\tau$-pair production rate in the low background environment of $e^+e^-$ -collisions at 10.58~GeV. 
Up to mid-2021 Belle~II collected a sample corresponding to $214~\text{fb}^{-1}$ of data. 
We present a first measurement of the $\tau$ mass, the prospects for the $\tau$-lifetime measurement, and review the overall $\tau$ lepton physics program of Belle~II.}
\FullConference{%
	  *** 10th International Workshop on Charm Physics (CHARM2020), ***\\
	*** 31 May - 4 June, 2021 ***\\
	 *** Mexico City, Mexico - Online ***
	}
\begin{document}
\maketitle

\section{Introduction}
SuperKEKB~\cite{AKAI2018188} is an energy-asymmetric $e^+e^-$-collider located in Tsukuba, Japan. It operates at the $\Upsilon$(4S) resonance, which corresponds to a centre-of-mass energy of 10.58~GeV. 
It uses the nano-beam scheme~\cite{10.1093/ptep/pts083} to reach an expected instantaneous luminosity of about $6 \times 10^{35}~\text{cm}^{2}\text{s}^{-1}$. 
Although SuperKEKB is optimised for $B$-meson physics, it also provides an ideal environment for $\tau$-lepton physics. The cross-section for $\tau$-pair production is almost as high as the cross-section for $B$-meson production from the $\Upsilon$(4S) resonance, with about nine billion tau pairs per attobarn. Belle~II profits from a low background environment, a well-known initial state that constrains the $\tau$ kinematics, and efficient trigger systems.

\begin{figure}
	\centering
	\includegraphics[width=0.95\linewidth]{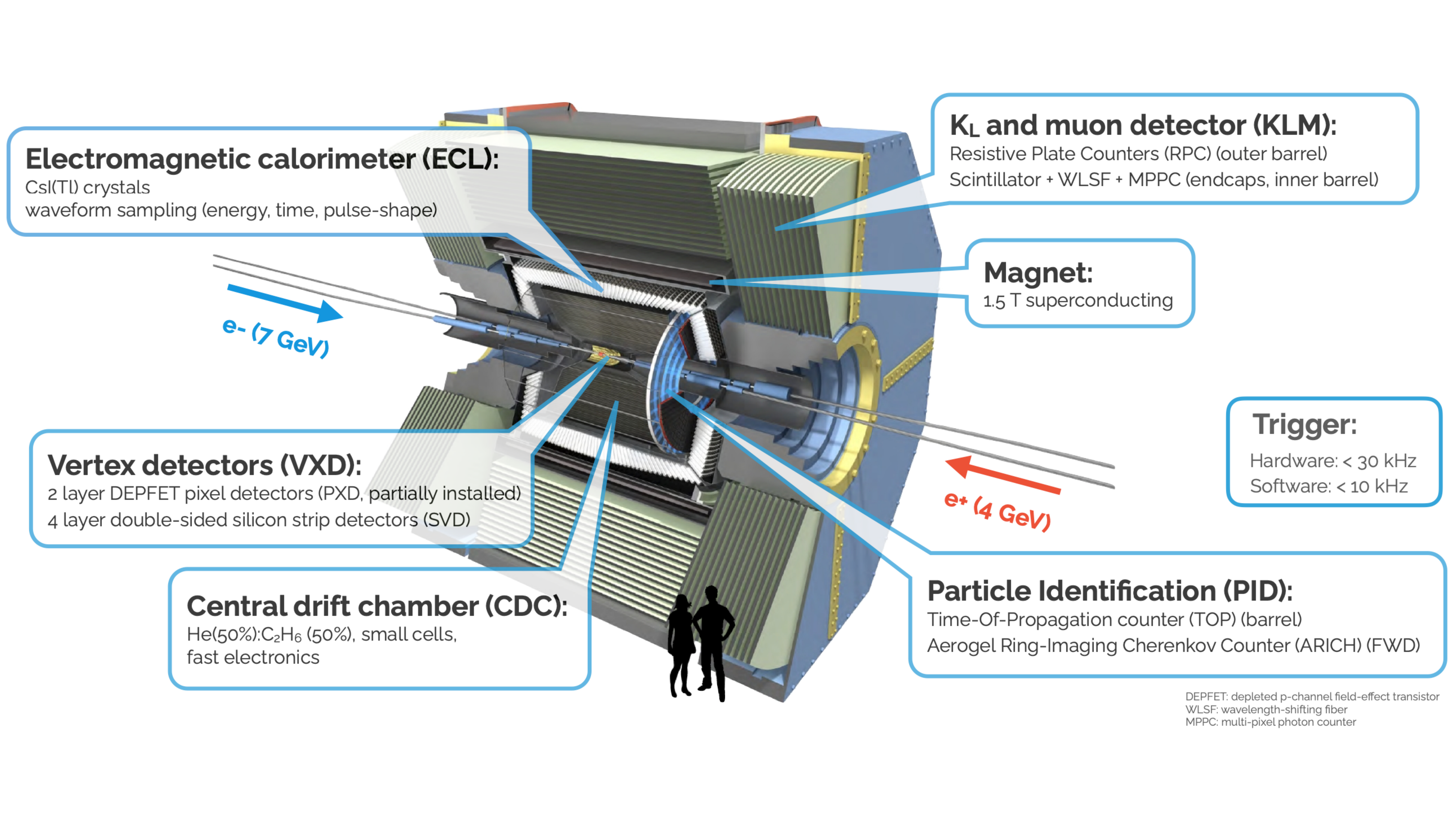}
	\caption{Schematic of the Belle II detector.}
	\label{fig:Belle2}
\end{figure}

The Belle~II detector surrounds the interaction point of SuperKEKB~\cite{Belle-II:2010dht}. It has the same general layout as its predecessor Belle, but with upgraded sub-detectors. Belle~II can cope with higher occupancies and data rates, and with the increased beam background rate due to the higher instantaneous luminosity of SuperKEKB compared to its predecessor KEKB. 
Central elements of the Belle~II upgrade are the vertex detector closer to the interaction point (IP), with the two innermost layers being pixel detectors and the remaining four silicon strip detectors. 
This improves the vertex resolution, compared to Belle, by almost a factor of two. 
The replaced central drift chamber has a larger volume with smaller drift cells, which improves the charge-particle momentum resolution. Also, a completely new particle identification system was installed, with a time-of-propagation detector in the barrel, and an aerogel-ring imaging Cherenkov counter in the forward endcap.
Figure~\ref{fig:Belle2} shows the full layout of the Belle~II detector.

Belle~II aims to collect a total of $50~\text{ab}^{-1}$, which corresponds to about forty-five billion $\tau$-pair events.
In the following, we discuss the expected increase in precision in  $\tau$-property measurements, such as the $\tau$ mass or $\tau$ lifetime, and the potential  to probe the mass scale or coupling strength for beyond the Standard Model physics with unprecedented depth. Especially in the search for lepton flavour violation we expect to improve the sensitivity by about two orders of magnitude. The quality of early Belle~II data is illustrated by a first measurement of the $\tau$-mass.

\section{Measurements of $\boldsymbol{\tau}$ Properties}

The precise measurement of $\tau$-properties is important for Standard Model predictions and for beyond the Standard Model physics searches. 
Furthermore, there are parameters, such as the electric dipole moment, $d_{\tau}$, or the anomalous magnetic moment, $a_{\tau}=(g-2)/2$, with measurements consistent with zero. 
Belle~II has the potential to further constrain Standard Model deviations and for the fist non-zero measurement of $a_{\tau}$.

\subsection{$\boldsymbol{\tau}$-Mass Measurement}
\begin{figure}
	\centering
	\subfloat[Pseudomass distribution of Belle~II data, corresponding to $8.8~\text{fb}^{-1}$, with fit projection overlaid
	]{
		\begin{minipage}{0.5\linewidth}
			\includegraphics[width=0.95\linewidth]{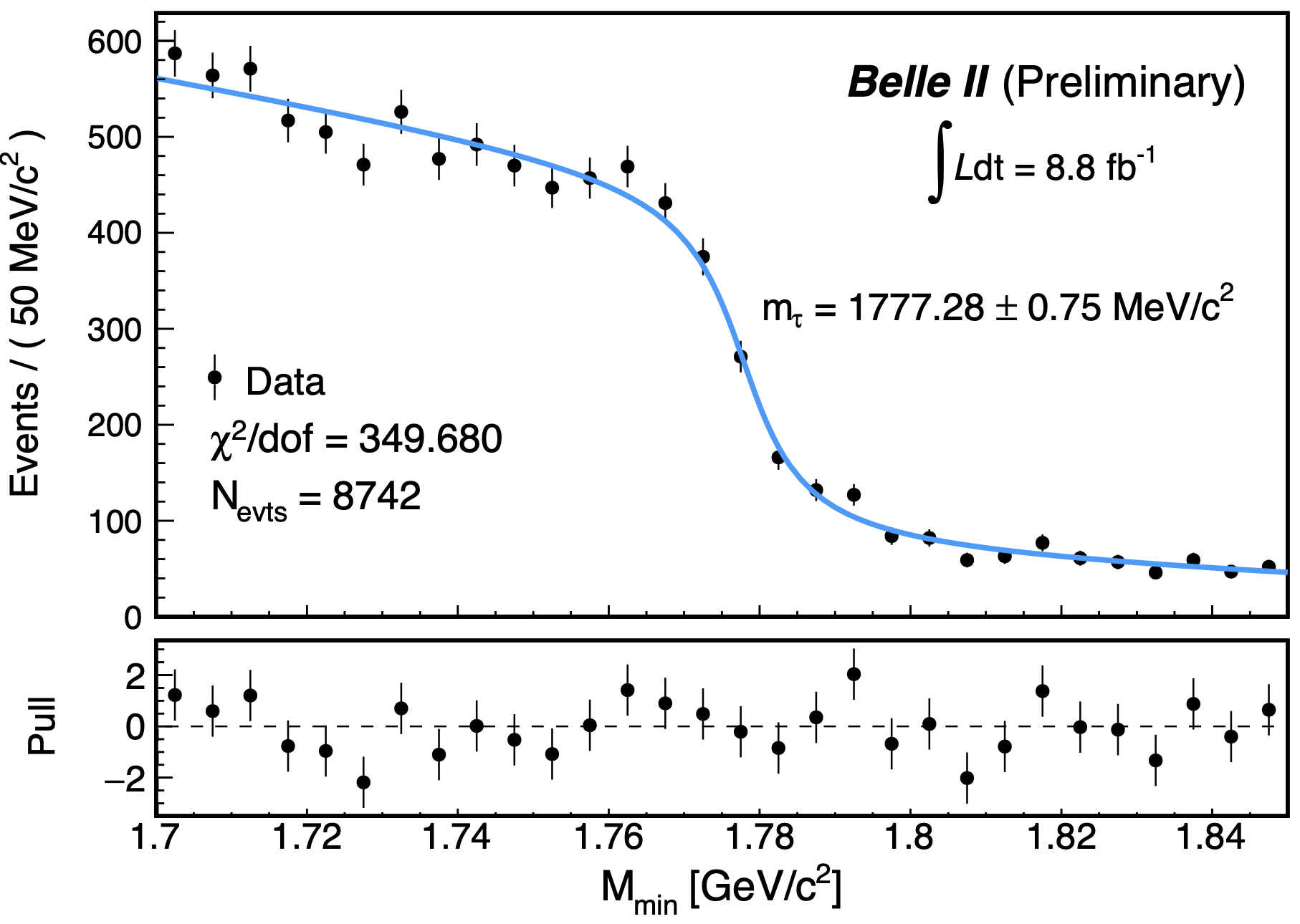}
			\label{fig:tauMass1}
		\end{minipage}
	}
	\subfloat[Comparison of $\tau$-mass measurements.]{
		\begin{minipage}{0.5\linewidth}
			\includegraphics[width=0.95\linewidth]{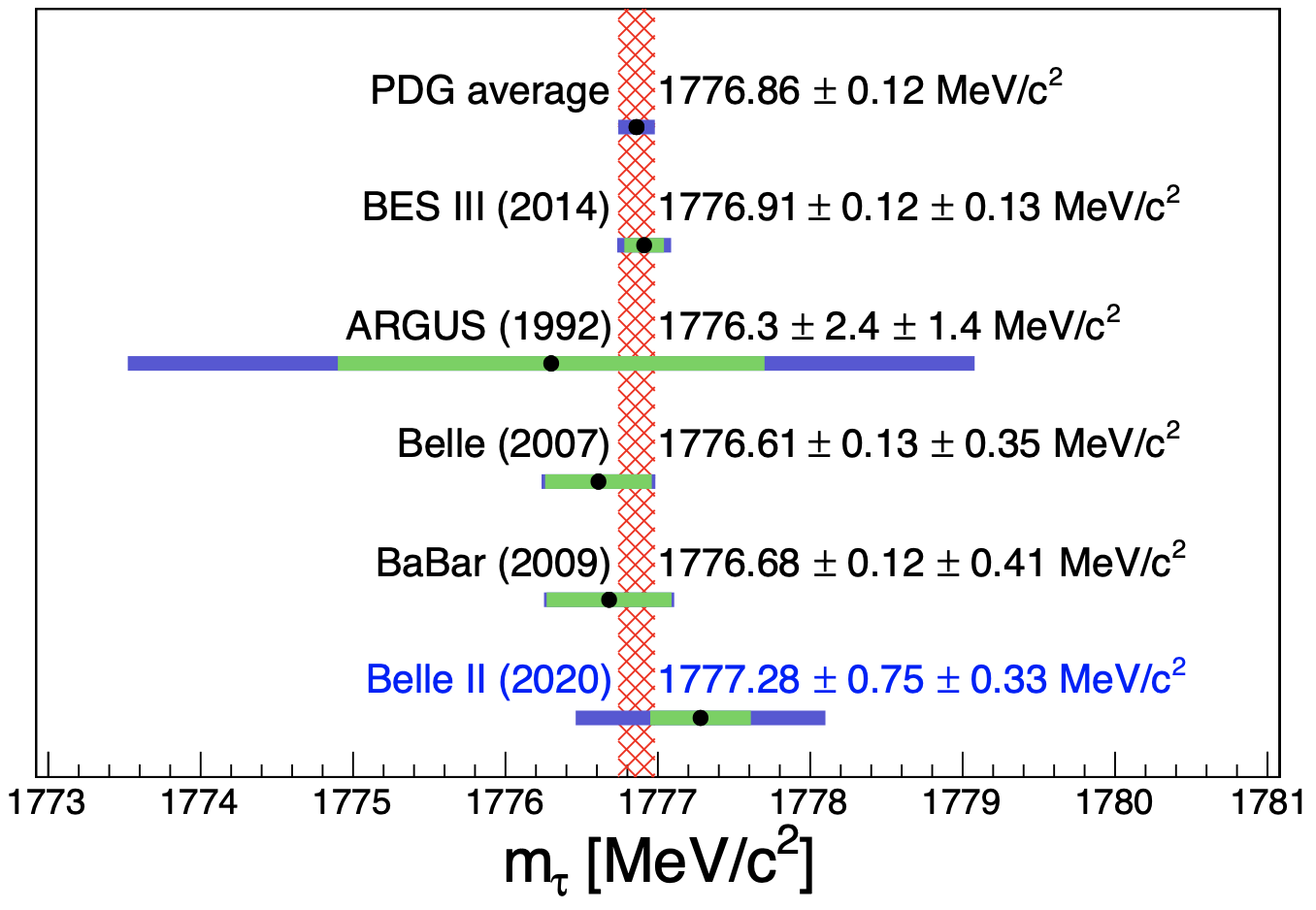}
			\label{fig:tauMass2}
		\end{minipage}
	}
	\caption{$\tau$-mass measurement of Belle~II using the pseudo mass technique from ARGUS. 
				}
\end{figure}

Belle~II follows the approach of ARGUS~\cite{ARGUS:1992chv} to measure the $\tau$ mass, $m_{\tau}$. Here, the pseudomass of the decay channel $\tau^- \to \pi^-\pi^+\pi^- \nu_{\tau}$ is used
\begin{equation}
	M_{\min} = \sqrt{m_{3\pi}^2 + \frac{2}{c^4}(\frac{\sqrt{s}}{2} - E_{3 \pi})(E_{3 \pi} - cp_{3 \pi})},
\end{equation}
where $m_{3\pi}$ is the mass of the 3$\pi$-system, $\sqrt{s}$ is the centre of mass energy, $c$ is the speed of light in a vacuum set to one in natural units from now on, $E_{3\pi}$ is the energy of the 3$\pi$-system, and $p_{3 \pi}$ is the four-momentum of the 3$\pi$-system.
The measured $\tau$ mass, $$m_{\tau} = (1777.28 \pm 0.75 \pm 0.33)\,\mathrm{MeV/c^2},$$  is given by the endpoint of the distribution, which is determined by fitting Belle~II data corresponding to $8.8~\text{fb}^{-1}$  using an empiric fit function. For example, this function can be a modified sigmoid or arctangent, as shown in Figure~\ref{fig:tauMass1}. 

In this study, Belle~II focused on improving the systematic uncertainties. 
The main systematic uncertainty is the momentum scale factor followed by the estimator bias and the choice of the fit function. The combination of all evaluated systematic uncertainties cover smaller contributions for the moment, such as resolution smearing. 
The result is consistent with the average value reported by the Particle Data Group~\cite{Belle-II:2020wbx}, and has a similar systematic uncertainty as the measurements by Belle and BaBar. 
Belle~II is expected to reduce the systematic uncertainties further, thanks to improved understanding of the detector. 
The precision of the $\tau$ mass measurements at Belle~II is expected to surpass that of the $B$-factories with the currently available data set of slightly more that $200~\text{fb}^{-1}$.

\subsection{$\boldsymbol{\tau}$-Lifetime Measurement}
The currently most precise $\tau$ lifetime measurement was obtained by Belle, with an integrated luminosity of $711~\text{fb}^{-1}$, $\tau_{\tau} = 290.17 \pm 0.53 \pm 0.33~\text{fs}.$
Belle~II could improve this result with about $200~\text{fb}^{-1}$, owing to a nanometre scale beamspot and improved vertex resolution, which allows to include the more abundant 3x1 prong topology in addition to the 3x3 prong topology, used in the Belle analysis.

\begin{figure}
	\centering
	\subfloat[Sketch of interaction region.]{
		\begin{minipage}{0.51\linewidth}
			\includegraphics[width=0.95\linewidth]{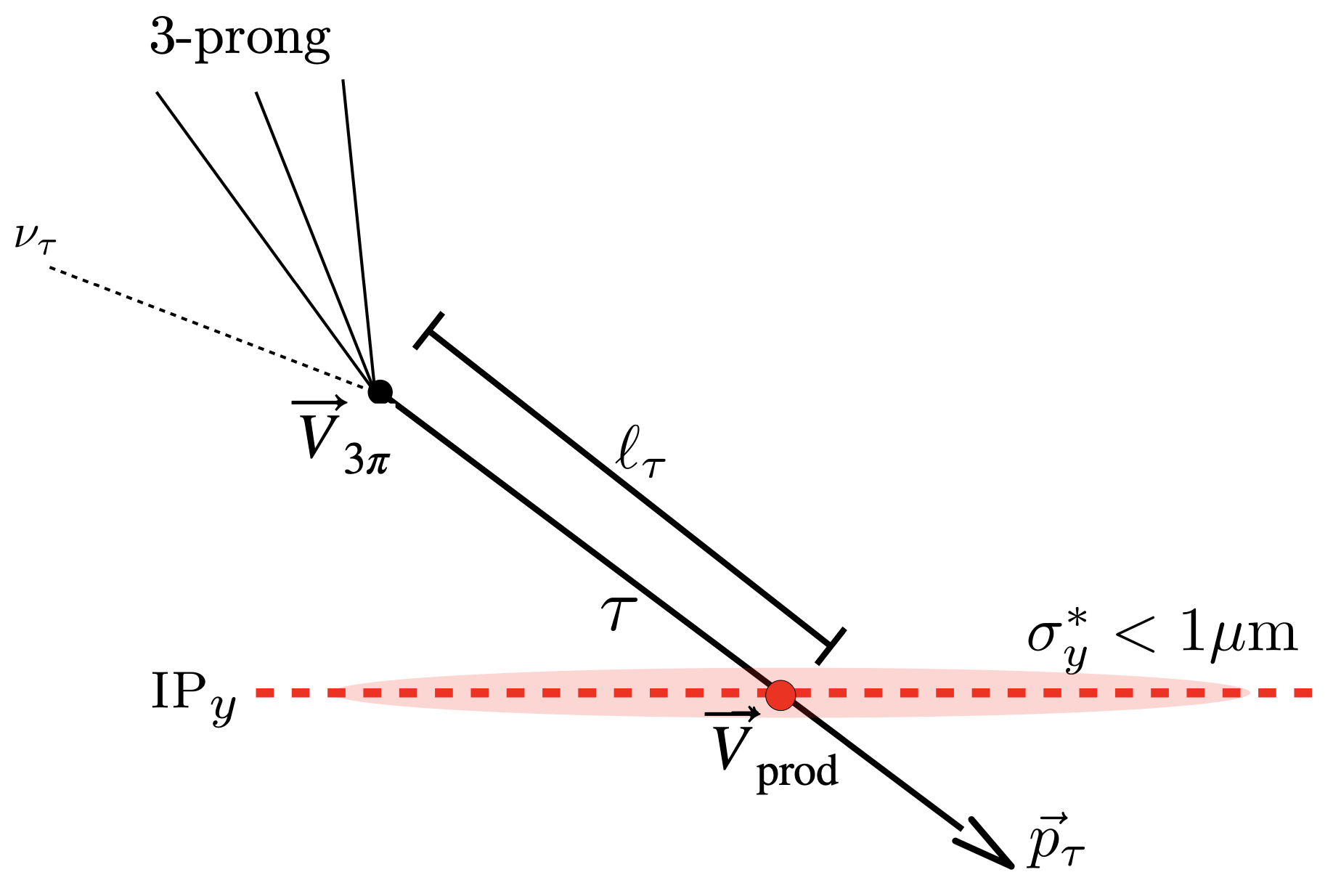}
			\label{fig:liftetimeScetch}
		\end{minipage}
	}
	\subfloat[Comparison of the measured $\tau$ decay-length resolution in Belle with simulated Belle~II data (black dots).]{
		\begin{minipage}{0.49\linewidth}
			\includegraphics[width=0.95\linewidth]{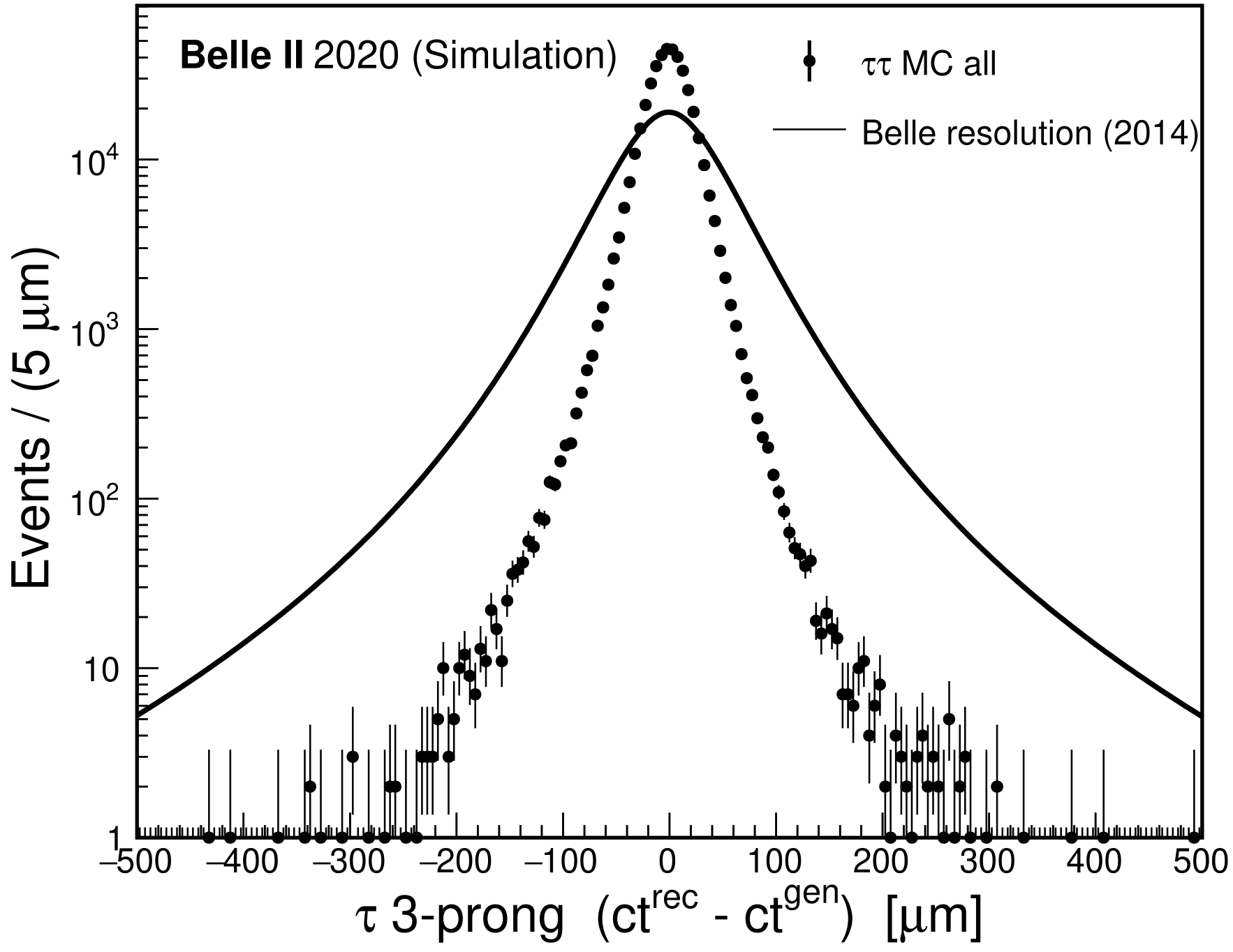}
			\label{fig:resulution}
		\end{minipage}
	}
	\caption{Sketch of the $\tau$-lifetime measurement principle (left) and expected resolution (right). 
		Due to the small beam size at the IP in $y$-direction, $\sigma_{y}^*$, and
		assuming that the $\tau$-pair production vertex, $\vec{V}_{\text{prod}}$, is resolved sufficiently by the size of the IP, the $\tau$ decay-length, $\ell_{\tau}$, is determined from the distance of only one $3\pi$-decay vertex, $\vec{V}_{3\pi}$, and the IP.
	} 
\end{figure}

The $\tau$ lifetime is determined from the decay time, $t_{\tau} = m_{\tau} \frac{\ell_{\tau}}{p_{\tau}}$, which is calculated from the observed decay length $\ell_{\tau}$ and the $\tau$ momentum, approximated by $p_{\tau} = \sqrt{\big(\frac{\sqrt{s}}{2}\big)^2 - m_{\tau}^2}$,
neglecting initial state radiation and final state radiation, as illustrated in Figure \ref{fig:liftetimeScetch}.
In Figure \ref{fig:resulution} we show the Belle~II decay time resolution in simulations, which is improved by almost a factor of two compared to Belle.
%
\subsection{Further Standard Model Measurements}
In addition to the mass and lifetime, Belle~II currently works on measurements of two notable quantities, $d_{\tau}$ and $a_{\tau}$. In light of the recent $g-2$ result of the muon~\cite{PhysRevLett.126.141801}, there is an increased interest in these parameters.
To date, Belle published the most precise measurement with about $30~\text{fb}^{-1}$ of data for the $d_{\tau}$~\cite{Belle:2002nla}.
An update using $833~\text{fb}^{-1}$, which improves approximately by a factor of five to an upper limit of $-1.85 \times 10^{-17} < \Re(d_{\tau}) < 0.61 \times 10^{-17}$ and $-1.03 \times 10^{-17} < \Im(d_{\tau}) < 0.23 \times 10^{-17},$ was recently submitted for publication~\cite{Belle:2021ybo}.

While the $d_{\tau}$ is beyond the precision achievable with the current Belle~II data set, the analysis will be able to provide a measurement of $a_{\tau}$ from the $\tau$-pair production vertex.
Today, the most precise measurement was performed with the process $e^+e^- \to e^+e^-\tau^+\tau^-$ by the DELPHI collaboration~\cite{DELPHI:2003nah}, which obtained a result consistent with zero and an uncertainty one order of magnitude larger than the Standard Model prediction, $a_{\tau} = 0.018 \pm 0.017$.
In the long-term, the full Belle~II data set may enable reaching a precision necessary~\cite{Belle2PhysBook} to probe the Standard Model prediction~\cite{Eidelman:2007sb} of $\frac{g-2}{2} \equiv a_{\tau}^{\rm \text{SM}} = (1.17721 \pm 0.00005)\times 10^{-3}$.

\section{Lepton Flavour Violation}
\begin{figure}
	\centering
	\includegraphics[width=0.95\linewidth]{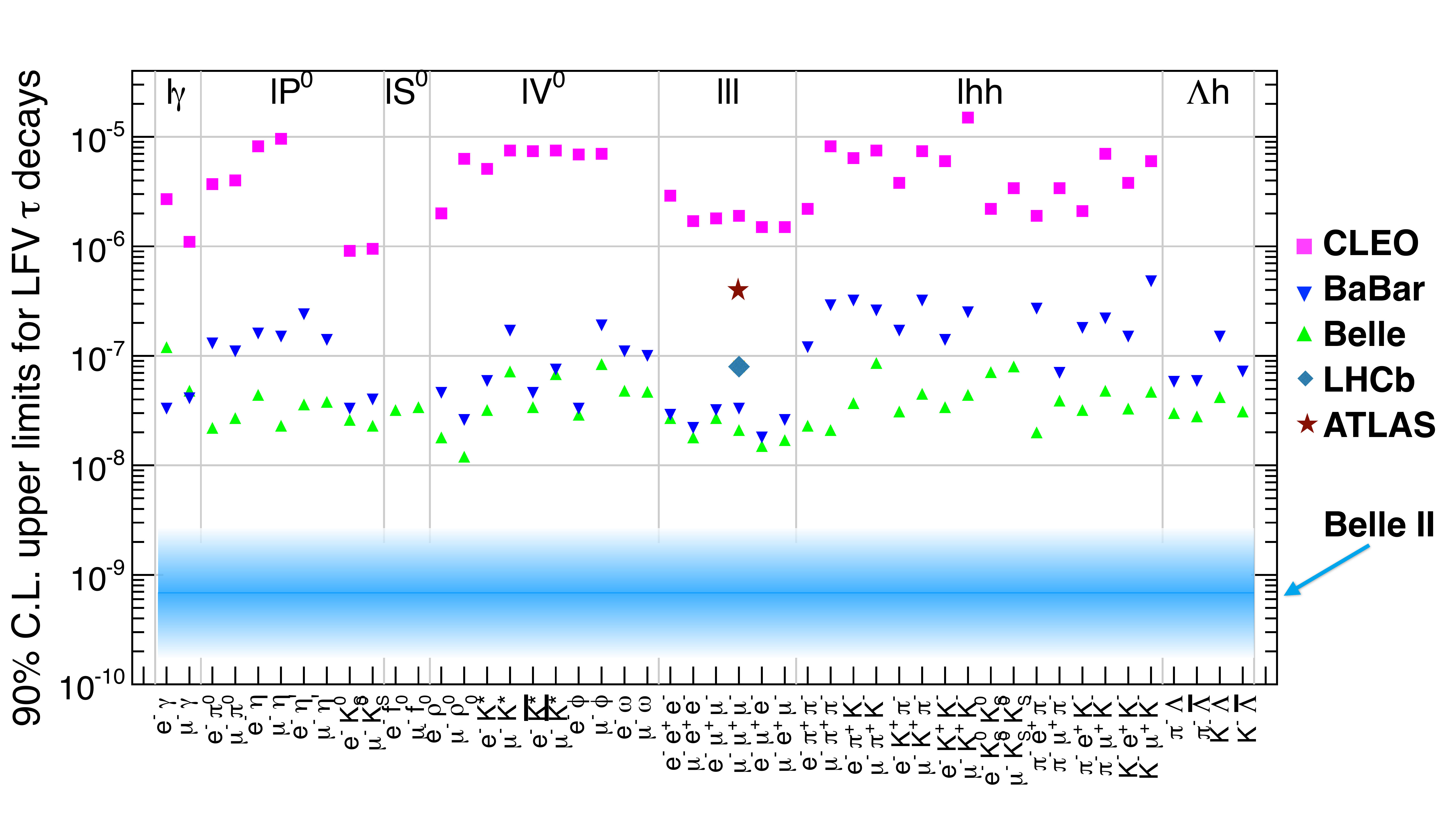}
	\caption{Comparison of sensitivity prospects for the $\tau$ lepton favour violating program at Belle~II with results obtained by other experiments.}
	\label{fig:LFVprogramm}
\end{figure}

In recent years lepton universality violation and lepton flavour violation have gained more and more interest due to the results of LHCb~\cite{Aaij:2021vac}. The Standard Model predicts no deviations from lepton universality and only highly suppressed lepton flavour violating processes, which are induced by neutrino oscillations. 
Figure \ref{fig:LFVprogramm} shows the expected sensitivity for all lepton flavour violating processes in the $\tau$ sector at Belle~II, which will improve current results by one order of magnitude or more.
In the following we discuss searches for lepton flavour violation in $\tau \to \mu \mu \mu$ and $\tau \to \ell \alpha$, where early results are expected.
\subsection{Early Results for $\boldsymbol{\tau \to \mu\mu\mu}$}
\begin{figure}
	\centering
	\includegraphics[width=0.5\linewidth]{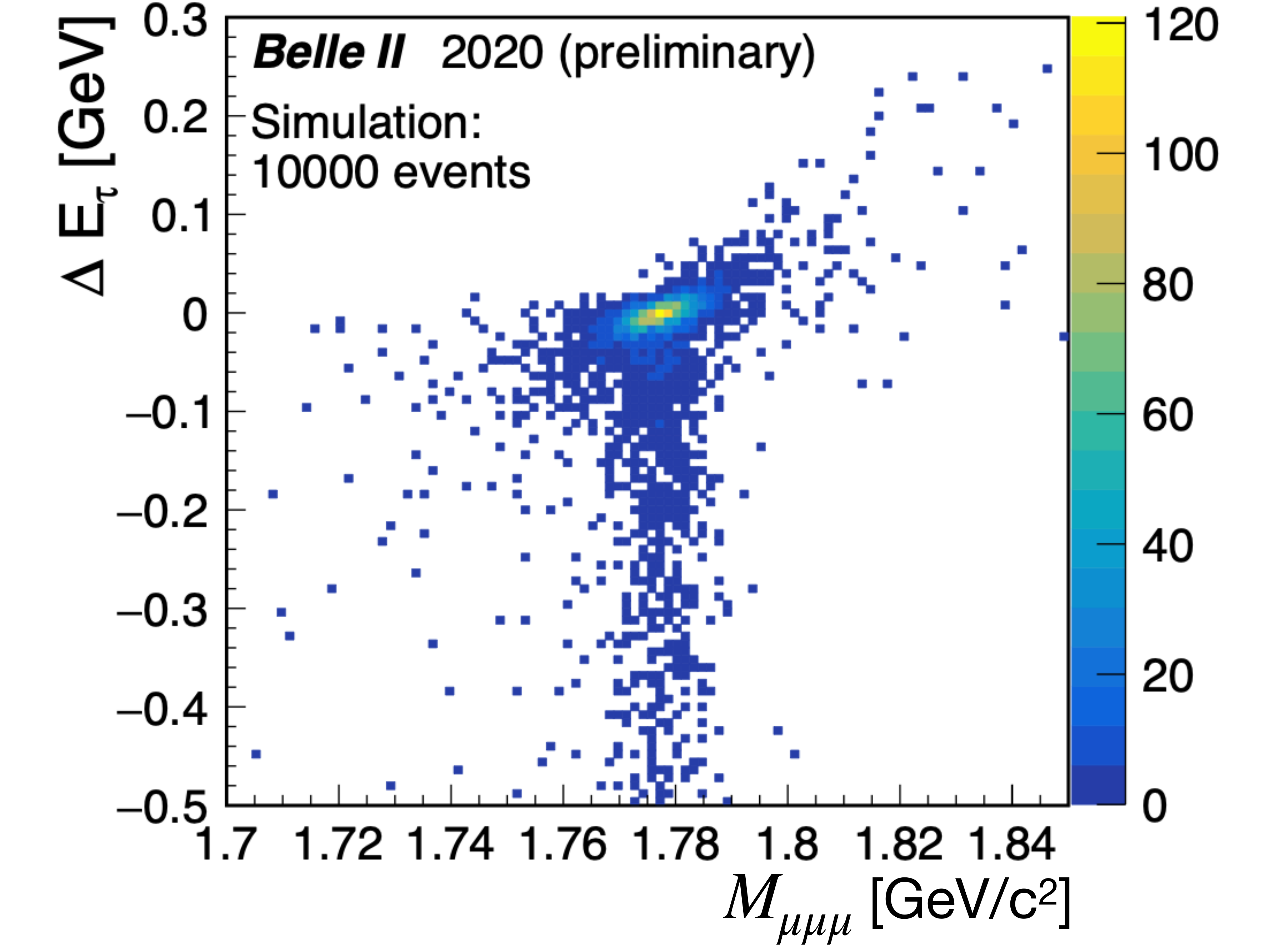}
	\caption{Expected distribution for the $\tau \to \mu \mu \mu$ decay in the missing energy ($\Delta E_{\tau}$) and reconstructed invariant mass ($M_{\tau}$) plane of the $\tau$-decay.}
	\label{fig:tau3mu}
\end{figure}

The search for the lepton-flavour-violating process $\tau \to \mu \mu \mu$ has a highly suppressed background because we expect the 3$\mu$ system of the $\tau$ to have an energy of $E_{\mu\mu\mu}=\sqrt{s}/2$ and an invariant mass  of $M_{\mu\mu\mu}=m_{\tau}$.
We expect a peak in the two-dimensional distribution of the missing energy of the $\tau$, $\Delta E_{\tau}= E_{\mu\mu\mu} - \sqrt{s}/2$, at zero and at $m_{\tau}$ for $M_{\mu\mu\mu}$, as illustrated  
in Figure \ref{fig:tau3mu}.

Because the background in this search is suppressed, the uncertainties should scale with sample size. Belle was able to set a limit at $Br(\tau \to \mu \mu \mu) < 2.1\times 10^{-8}$~\cite{Hayasaka:2010np}. To improve this result, Belle~II can either collect more data than the previous search, or try to improve the signal selection, which motivated a reassessment of the selection criteria used by Belle. 
By introducing a momentum dependent muon identification, increasing the muon momentum range, and allowing for the non-signal $\tau$, used to identify the $\tau$-pair event, to be also a $\tau \to \mu \nu_{\tau} \nu_{\mu}$ the signal selection efficiency is improved by approximately a factor of two. 
With these changes and an overall improved analysis competitive results are expected to be achieved with about $400~\text{fb}^{-1}$ of data.

\subsection{Early Results for $\boldsymbol{\tau \to \ell \alpha}$}
Belle~II pursues a generic search for a beyond the Standard Model, invisible particle $\alpha$ using the lepton flavour violating decay $\tau \to \ell \alpha$. 
There are several theories that propose $\tau \to \ell \alpha$, including some $Z^{\prime}$~\cite{ALTMANNSHOFER2016389} and axion-like particle models~\cite{calibbi2020looking}. The $\tau \to \ell \alpha$ process is especially sensitive to probe $Z^{\prime}$ models. In the case of the axion-like particles, $\tau \to \ell \alpha$ has a unique parameter space above the $\mu$ mass.

\begin{figure}
	\centering
	\subfloat[
	]{
		\begin{minipage}{0.47\linewidth}
			\includegraphics[width=0.95\linewidth]{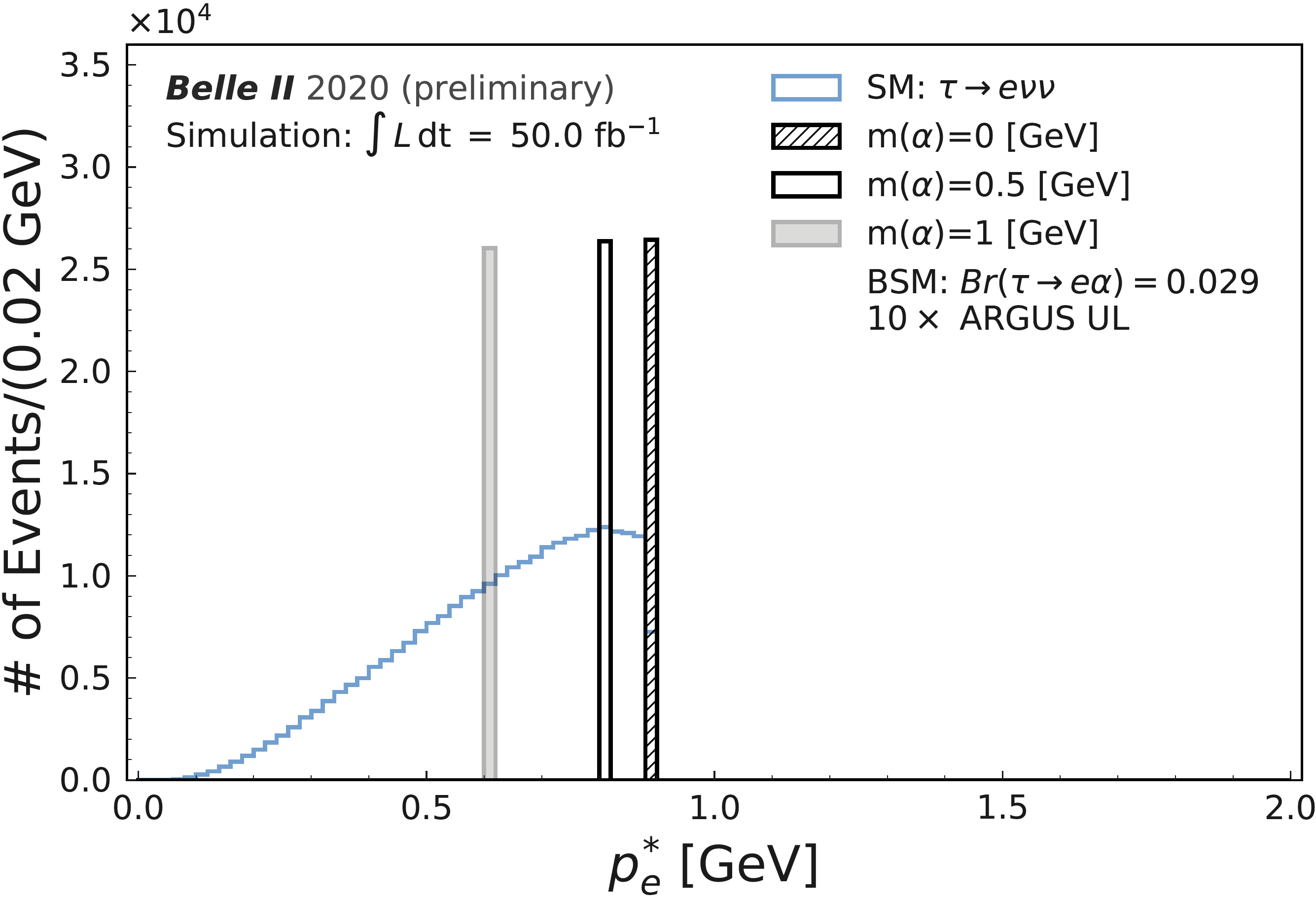}
			\label{fig:tauLa_gen}
		\end{minipage}
	}
	\subfloat[
	]{
		\begin{minipage}{0.53\linewidth}
			\includegraphics[width=0.95\linewidth]{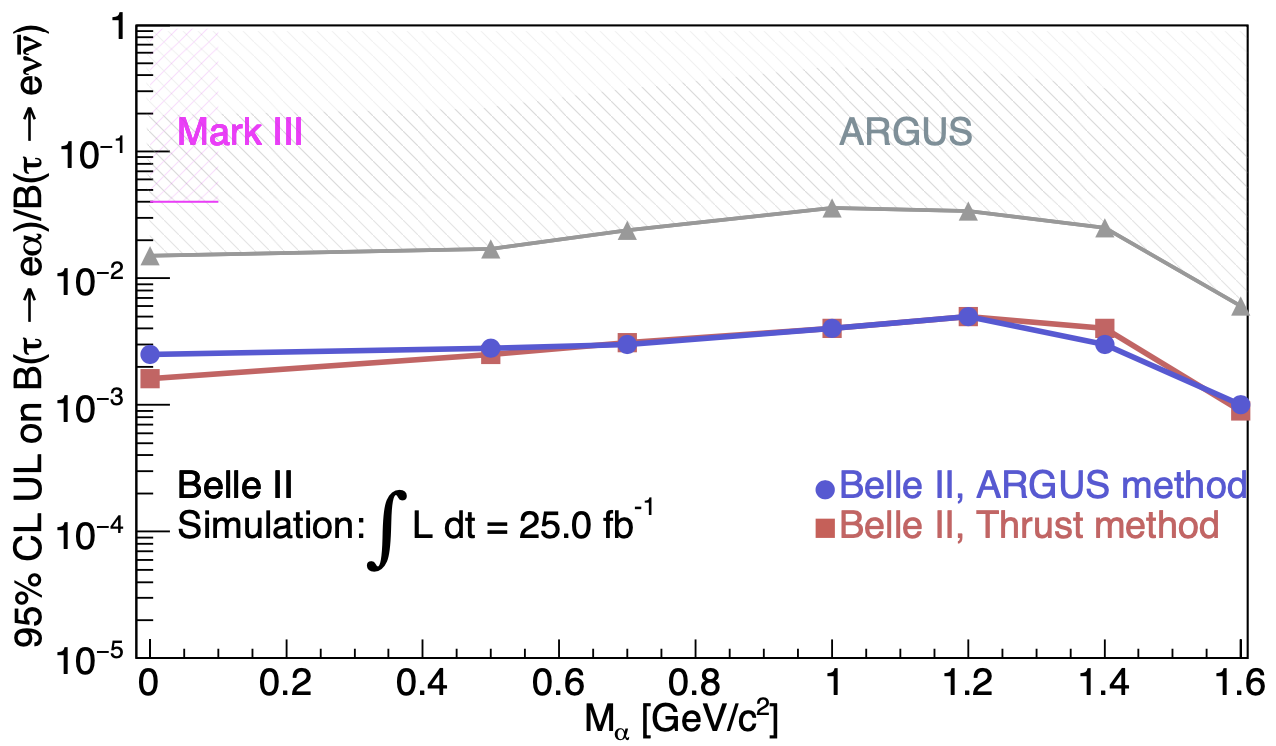}
			\label{fig:tauLaResulution}
		\end{minipage}
	}
	\caption{Expected $e$ momentum distribution for $\tau \to e \alpha$ compared to the Standard Model prediction in the $\tau$ rest-frame (left). Upper limit estimate as a function of the invisible mass for a simulation study without systematic uncertainties taken into account (right). }
\end{figure}

The search strategy is to look for a shape difference to the predicted Standard Model distribution. There is no unique signal region distinguishing between the Standard Model and the lepton flavour violating decay. 
The dominating background is the Standard Model decay into a lepton and two neutrinos. In the rest-frame of the $\tau$, the signal results in a mono-energetic lepton, while the background has a wide distribution given by the lepton decay parameters. 
Figure \ref{fig:tauLa_gen} shows the expected monochromatic $\ell$-momentum distribution for several mass hypotheses of $\tau \to \ell \alpha$ in the rest-frame of the $\tau$, compared to the Standard Model background.

The rest-frame estimate relies on the tag-$\tau$, which is used to identify the event. Belle~II studied two methods for estimating the $\tau$ rest-frame. The ARGUS method~\cite{Albrecht:1995aa} infers the $\tau$ rest-frame by using the $\tau \to \pi \pi \pi \nu$ decay to estimate the $\tau$ momentum. Here, the $3\pi$-system estimates the direction of the $\tau$. The so-called thrust method replaces the momentum direction of the $3\pi$-system with the thrust vector, determined from all visible particles in the event, for the estimate of the $\tau$ direction.

Belle~II uses a limit-setting procedure based on the CLs method, which compares the likelihood of $\tau \to \ell \alpha$ + $\tau \to \ell \nu\nu$ with the likelihood of $\tau \to \ell \nu\nu$ only. Here, a template fit to the lepton momentum in the rest-frame of the signal $\tau$ determines the likelihood from the best fit.
The template fit is based on three template distributions that describe the data
\begin{equation}
	f(x) = N_{ \ell \alpha } \cdot f_{\ell \alpha}(x) + N_{\ell \nu \nu} \cdot f_{\ell \nu \nu}(x) + N_{\text{BG}}\cdot f_{\text{BG}}(x),
\end{equation}
where $N_{\ell \alpha}$ is the signal yield, $N_{\ell \nu \nu}$ the yield of the $\tau \to \ell \nu \nu$ process, $N_{\text{BG}}$ the background yield; $f_{\ell \alpha}(x)$ the signal probability density function (pdf), $f_{\ell \nu \nu}(x)$ pdf of the $\tau \to \ell \nu \nu$ process, and $f_{\text{BG}}(x)$ background pdf.
After the selection and the fit, if no signal is present, an upper limit is obtained. 
The upper limit expected for an integrated luminosity of $25~\text{fb}^{-1}$ is shown in Figure~\ref{fig:tauLaResulution} for both rest-frame techniques. 

%

\section{Conclusion}
Belle~II is currently working on improving two Standard Model parameters, the $\tau$ mass and $\tau$ lifetime. 
In the mid- to long-term, Belle~II has the potential to measure the $d_{\tau}$ and $a_{\tau}$ of the $\tau$, to improve the precision of other Standard Model parameters, and for improving the sensitivity to lepton flavour violating processes in many possible decay channels, with early competitive results expected in $\tau \to \mu \gamma$ and $\tau \to \ell \ell \ell$.
Potential for increased sensitivity is already evident in the early phase in the lepton flavour violating sector for the $\tau \to \mu \mu\mu$ and $\tau \to \ell \alpha$ decays. 
Belle~II will lead the study of the $\tau$ physics sector for at least the next decade. 
\bibliographystyle{unsrt}
\bibliography{/Users/thomas/Dropbox/Doktor/Dissertation/DissertationBibTex.bib}

\begin{thebibliography}{10}

\bibitem{AKAI2018188}
Kazunori Akai et~al.
\newblock Superkekb collider.
\newblock {\em Nuclear Instruments and Methods in Physics Research Section A:
  Accelerators, Spectrometers, Detectors and Associated Equipment},
  907:188--199, 2018.
\newblock Advances in Instrumentation and Experimental Methods (Special Issue
  in Honour of Kai Siegbahn).

\bibitem{10.1093/ptep/pts083}
Yukiyoshi Ohnishi et~al.
\newblock {Accelerator design at SuperKEKB}.
\newblock {\em Progress of Theoretical and Experimental Physics}, 2013(3), 03
  2013.
\newblock 03A011.

\bibitem{Belle-II:2010dht}
T.~Abe et~al.
\newblock {Belle II Technical Design Report}.
\newblock {\em arXiv}, (1011.0352), 2010.

\bibitem{ARGUS:1992chv}
H.~Albrecht et~al.
\newblock {A Measurement of the tau mass}.
\newblock {\em Phys. Lett. B}, 292:221--228, 1992.

\bibitem{Belle-II:2020wbx}
F.~Abudin\'en et~al.
\newblock {$\tau$ lepton mass measurement at Belle II}.
\newblock {\em arXiv}, (2008.04665), 2020.

\bibitem{PhysRevLett.126.141801}
B.~Abi et~al.
\newblock Measurement of the positive muon anomalous magnetic moment to 0.46
  ppm.
\newblock {\em Phys. Rev. Lett.}, 126:141801, Apr 2021.

\bibitem{Belle:2002nla}
K.~Inami et~al.
\newblock {Search for the electric dipole moment of the tau lepton}.
\newblock {\em Phys. Lett. B}, 551:16--26, 2003.

\bibitem{Belle:2021ybo}
K.~Inami et~al.
\newblock {An improved search for the electric dipole moment of the $\tau$
  lepton}.
\newblock {\em arXiv}, (2108.11543), 2021.

\bibitem{DELPHI:2003nah}
J.~Abdallah et~al.
\newblock {Study of tau-pair production in photon-photon collisions at LEP and
  limits on the anomalous electromagnetic moments of the tau lepton}.
\newblock {\em Eur. Phys. J. C}, 35:159--170, 2004.

\bibitem{Belle2PhysBook}
W.~Altmannshofer et~al.
\newblock {The Belle II Physics Book}.
\newblock {\em PTEP}, 2019(12):123C01, 2019.
\newblock [Erratum: PTEP 2020, 029201 (2020)].

\bibitem{Eidelman:2007sb}
S.~Eidelman and M.~Passera.
\newblock {Theory of the tau lepton anomalous magnetic moment}.
\newblock {\em Mod. Phys. Lett. A}, 22:159--179, 2007.

\bibitem{Aaij:2021vac}
Roel Aaij et~al.
\newblock {Test of lepton universality in beauty-quark decays}.
\newblock {\em arXiv}, (2103.11769), 3 2021.

\bibitem{Hayasaka:2010np}
K.~Hayasaka et~al.
\newblock {Search for Lepton Flavor Violating Tau Decays into Three Leptons
  with 719 Million Produced Tau+Tau- Pairs}.
\newblock {\em Phys. Lett. B}, 687:139--143, 2010.

\bibitem{ALTMANNSHOFER2016389}
Wolfgang Altmannshofer, Chien-Yi Chen, P.S.~Bhupal Dev, and Amarjit Soni.
\newblock Lepton flavor violating z-prime explanation of the muon anomalous
  magnetic moment.
\newblock {\em Physics Letters B}, 762:389 -- 398, 2016.

\bibitem{calibbi2020looking}
Lorenzo Calibbi, Diego Redigolo, Robert Ziegler, and Jure Zupan.
\newblock {Looking forward to Lepton-flavor-violating ALPs}.
\newblock {\em JHEP}, 09:173, 2021.

\bibitem{Albrecht:1995aa}
H.~Albrecht et~al.
\newblock A search for the lepton-flavour violating decays $\tau \to e \alpha$,
  $\tau \to \mu \alpha$.
\newblock {\em Z. Phys. C - Particles and Fields}, 68(1):25--28, 1995.

\end{thebibliography}

\end{document}